\documentstyle[prb,aps,twocolumn,epsf]{revtex}
\begin{document}
\draft
\preprint{\today}
\title{Competition between Magnetic and Structural Transitions in CrN}
\author{A. Filippetti, W. E. Pickett and B. M. Klein}
\address{Department of Physics, University of California -- Davis, 
Davis, Ca 95616}
 
\maketitle
%
\begin{abstract}

CrN is observed to undergo a paramagnetic to 
antiferromagnetic transition with N\'eel temperature T$_N$ $\sim$ 280 K, 
accompanied by a shear distortion from cubic NaCl-type to orthorhombic Pnma 
structure. Our first-principles plane wave calculations, based on
ultrasoft pseudopotentials, confirm that
the distorted antiferromagnetic phase with spin configuration arranged 
in double ferromagnetic sheets along the [110] direction is energetically
much more stable than the paramagnetic phase, and also slightly favored
over the ferromagnetic and other examined antiferromagnetic phases.  
The energy gain from polarization is much larger than that from distortion; 
nevertheless, the distortion is decisive in resolving the competition among
the antiferromagnetic phases: the anisotropy in the (100) plane arising
from the ferromagnetic double-sheet magnetic order allows a small but
important energy gain upon the cubic-to-orthorhombic transition. 
Although antiferromagnetic order leads to a large depletion of states 
around Fermi level, it does not open a gap. The system 
is metallic with occupied hole and electron pockets of Fermi 
surface containing $\sim 0.025$ carriers of each sign per formula unit.
The simultaneous occurrence of structural distortion and antiferromagnetic 
order, as well as the competition between antiferromagnetism and 
ferromagnetism, is analyzed.

\end{abstract}
\pacs{71., 75., 71.15.Hx, 75.50.Ee}

\narrowtext
 
\section{introduction}
$3d$ transition metal mono-nitrides (TMN) are interesting technologically
due to a rich variety of properties,
ranging from high-T$_c$ superconductivity (observed in TiN and VN),
to very hard surface coatings,
to heterogeneous catalysis\cite{intro}. Although some characteristics
are common to the whole series of compounds such as having high 
melting points and chemical hardnesses, this class also exhibits  
interesting variation in some properties.
ScN is semiconducting\cite{dyb}, as CrN is sometimes reported\cite{hhvp}
to be, while TiN and 
VN are superconductors \cite{gj}. Regarding magnetic properties, ScN, TiN and 
VN are paramagnets\cite{hhvp}, CoN has been only recently synthesized 
\cite{skymf} and is also found paramagnetic.  For CrN, strong experimental 
evidence supports antiferromagnetism below a N\'eel temperature 
T$\sim$273-286 K\cite{hhvp,ceh,ibcy,blsm,nesb}, 
and recent measurements suggest for FeN 
antiferromagnetism\cite{fen} as well. 

Even for the structural
properties there is no uniformity. Most of compounds 
have rocksalt structure, but there are exceptions: CoN has ZnS-type
structure\cite{skymf}, and FeN has been observed in both ZnS- and NaCl-type 
phases (depending on the growth conditions).  CrN seems to
be very peculiar in that it is the only one of the series that does not 
have a ground state with cubic symmetry.  
Its transition from paramagnetism
to antiferromagnetism at $T_N$ is 
accompanied by a structural distortion from cubic to orthorhombic (Pnma) 
symmetry\cite{ceh}.

The body of information on $3d$ TMN properties is rather fragmentary 
(and sometime contradictory), and much vital data are unavailable.
Most of these compounds are difficult to synthesize, and 
the strong tendency of nitrogen to form dimers leads to difficulty in
growing defect-free TMN crystals\cite{hhvp} (a problem also common to other 
nitride compounds such as III-V semiconductors). Thus far, nothing is known 
about MnN, NiN, CuN and ZnN, and only recently it has become possible to 
synthesize FeN\cite{fen} and CoN\cite{skymf}. 

Due to the difficulty in preparing stoichiometric CrN, and to a
strong sensitivity of its properties on the concentration of
defects, reports on the properties of CrN in the literature are
not entirely consistent.  Browne {\it et al.}\cite{blsm} have
studied the effects of variation of $x$ in CrN$_x$ and produced
samples with $x$ up to 0.997, taking care to keep them
oxygen-free.  Only at this high N composition is the
antiferromagnetic transition sharp, and it occurs at T$_N$=286 K.
The transformation is first order, with hysterisis of 2-3 K as
measured by linear expansivity, susceptibility, and resistivity.
At lower values of $x$ (down to 0.980) the transition
temperature decreases to 273 K and
the transition width increases.
 
At $x$=0.997, the properties reported by Browne {\it et al.}
are as follows.  The resistivity slope $d\rho/dT$ is
positive (metallic-like)
on both sides of the transition, and $\rho$ decreases by 30-35\%
across the transition, in opposition to the semiconducting behavior
reported by Herle {\it et al.}\cite{hhvp}  Due to the polycrystalline
nature, and specific density only 70\% of the bulk, the value of
the resistivity $\sim 2-3 m\Omega cm$ is only an upper bound.
The susceptibility slope changes from a small negative value
$d log\chi/dT \sim -4\times 10^{-4}$ K$^{-1}$ above T$_N$ to a
positive value of similar magnitude below T$_N$.  The decrease in
$\chi$ is $\sim$ 35\%.  These data are most consistent with a
metal-to-metal transition at T$_N$ in which a substantial fraction
of the Fermi surface is gapped by structural distortion and/or
the onset of magnetic order. Tsuchiya {\it et al.}\cite{films} have
produced thin films of CrN$_x$ with 1.0 $\leq$ $x$ $\leq$ 1.2.
These films, like the samples with $x$ $\leq$ 1 of Browne {\it et al.}, 
have T$_N$ $\leq$ 280 K. None of the transitions are as sharp as for the
$x$ = 0.997 samples. It is clear that properties are sensitive to N 
overstoichiometry as well as to understoichiometry.

Not surprisingly, the theoretical literature is also not very extensive.
Some work on the $3d$ TMNs in the cubic 
paramagnetic phase has been reported\cite{ppkb,nrews,zgjca,hg}, 
but very little is reported 
for compounds that undergo reconstructions accompanied by, and
perhaps driven by,
magnetic ordering, as may be the case in CrN\cite{mmsd} and FeN\cite{sss}. 
Investigation of compounds that are so close in chemical composition
but with so different physical properties are important in 
providing insight into basic mechanisms such as magnetic ordering, 
formation of local moments, and relations of this phenomena with basic 
features such as structure and density of states. Ab initio calculations,
which can separate the effects of magnetism and structural distortions,
are very effective in such cases. In this paper we focus on CrN 
and investigate the occurrence of magnetic ordering for the cubic structure
and for the (observed) orthorhombic structure.

A basic characteristic of CrN in the cubic paramagnetic phase, which
makes it easy to understand
the instability towards ferromagnetism, is the very high
density of states \cite{ppkb} at the Fermi energy (E$_F$). 
Less easy to understand is the
competition between AFM and FM ordering, and the relation between
spin polarization, magnetic order, and structural distortion.
Other cases are well known where high DOS drives systems to structural
(e.g. VO$_2$, from tetragonal to monoclinic) or spin polarized (CrO$_2$,
from PM to half-metallic FM) transitions. In fact, the isovalent compound 
MoN shows an interesting parallel.  Due to the peak at E$_F$ in the density 
of states in the cubic phase, it was predicted to be a high temperature 
superconductor\cite{ppkb2} (compared to conventional superconductors). 
It was subsequently shown by calculation\cite{cbkm} that it is structurally 
unstable in the cubic structure,
which explains the fact that experimentalists had been unable to grow
stoichiometric samples.  In the case of CrN the cubic structure is
stable at high temperature, but at T$_N$
there is the simultaneous occurrence of both
structural transformation and magnetic order.
Giving insight into possible relations between these transitions,
and identifying the driving force for the transition,
is the main motivation of the present work.

The paper is organized as follows: in Section \ref{csmc} a review of the
CrN crystal structure is given, together with a brief description of
our calculational procedure. In Section \ref{sp} we show structural and
energetic results for CrN in different magnetic phases. Finally, Section
\ref{ep} is dedicated to an analysis of the electronic properties. 
Results for density of states, band energies, and some Fermi surface 
properties are presented and discussed.

\section{Crystal Structure and Method of Calculation}
\label{csmc}

\begin{figure}
\epsfxsize=7cm
\epsffile{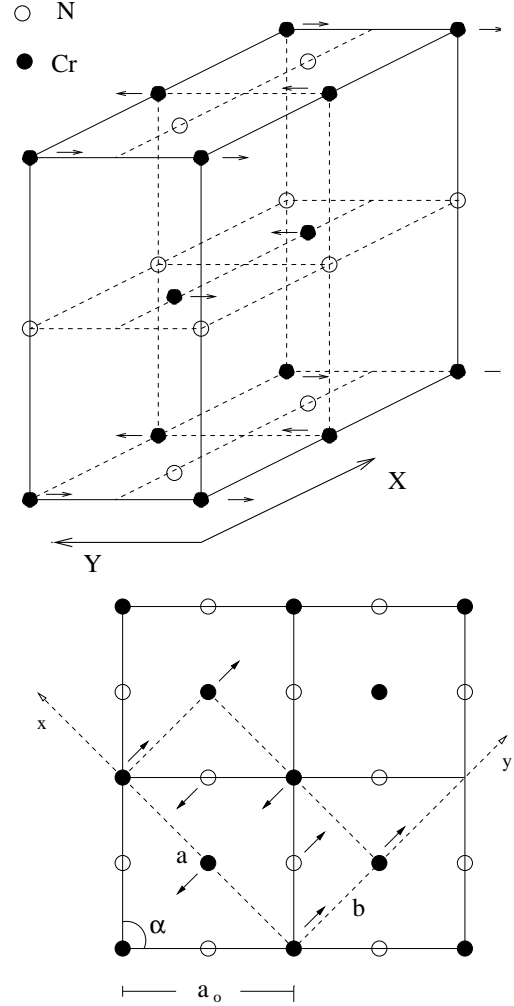}
\caption{
Top panel: perspective view of the CrN orthorhombic cell with
antiferromagnetic spin arrangement consisting of alternating
double ferromagnetic layers parallel to the [110] (following
Corliss {\it et al.}).
Arrows denote the relative spin directions on Cr atoms in the N\'eel state.
Bottom panel: Top view of (001) plane. The dashed line indicates the
orthorhombic cell with respect to the cubic structure.}
\label{fig1}
\end{figure}

In Figure \ref{fig1} a tridimensional perspective
(top panel) and a top view down onto the (001) plane (bottom panel)
of the CrN structure is shown. The $\hat{x}$, $\hat{y}$ and 
$\hat{z}$ axes of the unit cell for the AFM phase are oriented along 
the [1$\overline{1}$0], [110] and [001] directions of the cubic NaCl 
structure, respectively. The shear distortion consists of a 
contraction (not visible in the Figure) of the angle $\alpha$ from the 
cubic ($\alpha$= 90$^o$) to a $\sim$2$^o$ smaller observed value. 
The square symmetry in the (001) plane is broken so that the coordination 
of the atoms reduces to 2 in the (001) plane. Orthorhombic and cubic 
structures are connected by the relations $a=2a_0 sin(\alpha/2)$, 
$b=a_0 cos(\alpha/2)$ and $c= a_0$. 

From experiments, two possible antiferromagnetic spin 
alignments have been proposed\cite{ceh,nesb}. We consider here primarily 
the arrangement determined by Corliss {\it et al.}\cite{ceh} and depicted by 
arrows on Cr sites in Figure \ref{fig1}: double ferromagnetic 
layers parallel to the [110] direction alternate in such a way to form a 
four-layer antiferromagnetic orthorhombic unit cell (4 formula units and 
8 atoms per cell), with non vanishing magnetization field planarly averaged 
along [1$\overline{1}$0]. 
As a result of this ordering, six of the 12 (cubic) Cr neighbors have
spin parallel and six spin antiparallel. Of the six (cubic) next nearest
neighbors, four have antiparallel spin.
Ref. \onlinecite{ceh} reports for the magnetic moments $m$ = 2.36 $\mu_B$ 
per Cr atom.  Another experiment\cite{ibcy} gives a much larger $m$ = 3.17 
$\mu_B$ that seems to be much less reasonable. Indeed, the first value is 
consistent with a Cr to N charge transfer of $\simeq$ 2 electrons, while the 
second would imply an almost completely ionized configuration
Cr$^{+(3-\delta)}$N$^{-(3-\delta)}$ that is not to be expected (charge
transfer would be larger than in more ionic compounds such as transition-metal
oxides).

Another motivation for this work is technical and centers on our 
method of calculation.
These local-spin-density calculations are performed within a plane 
wave and ultrasoft pseudopotential\cite{van} (USPP) framework. Here
we want to furnish a sound assessment of accuracy and efficiency 
of the use of pseudopotentials for the study of materials historically 
considered impossible, or at best difficult, to be 
treated within a plane wave pseudopotential method. CrN is a difficult 
test, being the combination of a $3d$ transition metal and a first-row atom, 
with high density of states at E$_F$ and a reconstruction accompanied by
antiferromagnetism.

To our knowledge this is the first plane-wave study of AFM CrN (only one 
augmented-spherical-wave calculation\cite{mmsd} is present in literature).
The use of pseudopotentials to study magnetic compounds itself is only 
two years old\cite{las}. Use of USPP for magnetic materials is very 
new\cite{moroni,sawada} and presents the essential feature: constructing 
smooth (but well transferable) USPP allows us to obtain converged results 
(within 1 mRy for the total energy) at energy cut-off of 29 Ry 
($\sim$2600 plane waves per CrN unit), 
whereas ordinary 
norm-conserving pseudopotentials would require at least the twice 
as large a cutoff value, hence four times as many plane waves. 
The chosen core radii are 2.20 a.u., 2.45 a.u. and 2.45 a.u. for s, p and
d projectors of Cr pseudopotential, respectively; 1.30 a.u. for both s and p
components of N pseudopotential. Two projectors for each angular channel
(corresponding to energies $\sim$3 eV apart) were used.
The (highest in energy) p component was employed as local part.
 
The non-linear core correction (NLCC)\cite{lfc} was used to ensure the best
degree of transferability. As pointed out in Ref.\onlinecite{lfc},
the NLCC is particularly important for spin-polarized situations.
Indeed, pseudopotentials are generated in a non-spin-polarized
reference configuration. Such a procedure is justified only to the extent 
that the pseudopotential is perfectly descreened by the valence 
dependence. In case the overlap between core and valence charge is large, 
as it happens for transition metals, the non-linearity of the
exchange-correlation potential induces incomplete descreening, and
thereby a pseudopotential dependence on the valence configuration. Also,
the LSD exchange-correlation potential is even more non-linear than the
LDA one\cite{lfc}, and further affects the pseudopotential accuracy.
The NLCC set the pseudopotential free from its exchange-correlation
dependence, thus recovering its transferability.
Finally, no use of relativistic correction was made.
 
The performance of Cr and N pseudopotentials has been individually
tested. In particular, with the former we obtain an accurate description of
bcc bulk Cr (e.g. a calculated lattice constant within 0.5 \% the
experimental measure), whereas the latter has been estensively employed
in a long investigation of semiconducting III-V nitrides compounds
(see ref. \onlinecite{bfv} and references therein).
 
The exchange-correlation potential was that of Ceperley and Alder,
as parametrized by Perdew and Zunger\cite{vxc}.
We sampled the irreducible zone (IBZ) with special k-points
($\sim 30$ for self-consistent calculations, from 150 up to
530 k points to evaluate the densities of states).

\section{Structural Properties}
\label{sp}

In Table \ref{energy} our results for energies and magnetic moments of 
PM, FM and AFM phases of CrN are shown.
For a fair comparison between theory and experiment we investigated,
together with the AFM double-sheet ordering found by Corliss {\it et al} 
(here indicated as AFM$^2_{[110]}$), other two kinds of AFM orderings:
the single-sheet AFM$^1_{[110]}$ phase, similar to the AFM$^2_{[110]}$ 
but with single ferromagnetic sheets having alternated spin 
along [110] (a tetragonal structure with 4 atoms per cell), and the
AFM$_{[111]}$ phase, consisting on single ferromagnetic sheets parallel 
to the (111) plane which alternate the sign of the spin along [111]
(i.e. a trigonal structure common in $3d$ transition metal monoxides 
with 6 Cr and 6 N layers in the cell).
For each phase, we present results obtained in both cubic and distorted
structures. For the cubic structures we consider both experimental and
theoretical lattice constants (i.e. the $a_0$ obtained by energy minimization 
of the respective phases). To add the further effect of the planar distortion, 
we extract the value of $\alpha$ from the experimental structure given 
in Ref.\onlinecite{ceh} ($\alpha$=88.23$^o$).

A number of results for the PM cubic (NaCl) phase of CrN are present in 
literature\cite{mmsd,ppkb,sss}. 
In particular, Shimizu, Shirai, and Suzuki\cite{sss} performed lattice 
constant calculations for the whole series of 3d TMN's.  For CrN they 
report $a_0^{PM}$=7.56 a.u. which underestimates the experimental 
value $a_0^{exp}$=7.81 a.u. for the AFM$^2_{[110]}$ phase (a 10\% smaller 
volume). Our calculation gives $a_0^{PM}$=7.73 a.u., a value 1\% smaller 
than experiment. The discrepance between our $a_0^{PM}$ value and
that of Ref.\onlinecite{sss} can be ascribed to the different form of
the exchange-correlation used and, moreover to the fact that, differently
from Ref.\onlinecite{sss}, we perform a non-relativistic calculation.

For a true comparison with the experiments, however, it is necessary
to calculate the lattice constant for the cubic AFM$^2_{[110]}$ structure. 
We find $a_0^{AFM}$=7.876 a.u., less then 1\% above experiment. Thus, 
according to our calculations, polarization (from PM to AFM$^2_{[110]}$ order) 
produces an expansion of 2\% of the lattice constant. 

Looking at Table \ref{energy}, we see that the AFM$^2_{[110]}$ distorted 
phase is favoured overall. The theoretical determination of $a_0$ produces, 
with respect to the experimental structure, an energy gain 
(E$^{ex}_d$-E$^{th}_d$) of 0.02 eV/formula unit; anyway, in general, results 
obtained for theoretical
and experimental lattice constants are close each other to within 0.01 eV.
The spin polarization PM to AFM gives a considerable energy gain 
($\sim$ 0.15 eV to $\sim$ 0.3 eV per formula unit, depending on the 
magnetic phase), thus, polarization greatly lowers the energy regardless 
of the type of magnetic order. This is a sign that there is likely to 
be disordered 
local moment behavior above T$_N$, although it is not so evident from
the existing data.\cite{blsm}
Also, all AFM phases are more stable than the FM one.

Now, we examine in more detail the competition among the different 
AFM phases. The AFM$_{[111]}$ is worthwile to be considered because it 
represents the usual kind of AFM order observed for NaCl-type structures. 
Indeed, a peculiarity of the AFM$_{[111]}$ phase 
(unless one considers noncollinear arrangements) is having all the six 
(fcc) second neighbor moments antiparallel, a common characteristic of 
most of NaCl structures, as revealed by neutron diffraction studies\cite{nds}.
Nevertheless, we found the 
AFM$_{[111]}$ clearly disfavoured with respect to the AFM$^2_{[110]}$ and 
only slightly favoured on the FM phase. More interesting and somewhat
surprising is the result for the AFM$^1_{[110]}$ phase: we found it to 
be clearly the most stable of the cubic structures. It is characterized 
by four spin-parallel and eight spin-antiparallel first neighbors, and all 
the six second neighbors have parallel spins (an arrangement very different 
from the other AFM phases here considered). 
But if the planar distortion is added, the energetic order of AFM$^1_{[110]}$
and AFM$^2_{[110]}$ phases is reversed, and the AFM$^2_{[110]}$ becomes the 
lowest, thus recovering the agreement with the experiment.
Indeed, the distortion causes, for the AFM$^2_{[110]}$, the shortening of 
metal-metal distances between spin-antiparallel Cr atoms, that arguably
causes an energy gain, and the stretching of spin-parallel Cr atoms in the 
(100) plane (producing an energy loss), resulting, as a matter of fact,
in an appreciable net decrease of energy ($\sim$ 0.07 eV/formula unit). 
The same cannot occur for the AFM$^1_{[110]}$, because no spin-antiparallel 
first neighbors are in the (100) plane, and thus the distortion does not 
allow any stabilization (actually we found an energy increase of $\sim$ 0.01 
eV/formula unit).

Thus, the resulting physical picture is the following: if we consider the 
spin ordering in a constrained cubic symmetry, the AFM$^1_{[110]}$ phase,
having the largest number of fcc spin-antiparallel first-neighbors (8, 
against 6 of both AFM$^2_{[110]}$ and AFM$_{[111]}$ phases) is favoured. 
But if we allow the symmetry to relax, the (100) planar anisotropy of the 
AFM$^2_{[110]}$ phase causes some of the spin-antiparallel first neighbors 
to approach each other, producing a strong structural stabilization. 
We argue that in the cubic structure the bonds between spin-antiparallel 
metal atoms are under condition of large tensile stress (larger than that
in the bonds between spin-parallel atoms, anyway) 
and the cubic to orthorombic transition 
relieves part of this stress. Then, the planar distortion plays a key role
in the competition between AFM phases, and the stress relief is reasonably
to be considered the ultimate driving force towards distortion. 

For magnetic moments (we show in the Table only values referred to the
experimental structures, being very close to that obtained for the
theoretical ones) we obtain values ranging between $m$=1.9 $\mu_B$ and
2.3 $\mu_B$ per Cr atom. In particular, for the AFM$^2_{[110]}$ phase
our value $m$=2.15 $\mu_B$ is in fair agreement with the experimental
$m$=2.36 $\mu_B$ of Ref. \onlinecite{ceh}, and the theoretical
$m$=2.41 $\mu_B$ of Ref. \onlinecite{mmsd}.

\section{Electronic Properties}
\label{ep}

\subsection{Effect of AFM$^2_{[110]}$ Ordering}

\begin{figure}
\epsfysize=11cm
\epsfxsize=9cm
\epsffile{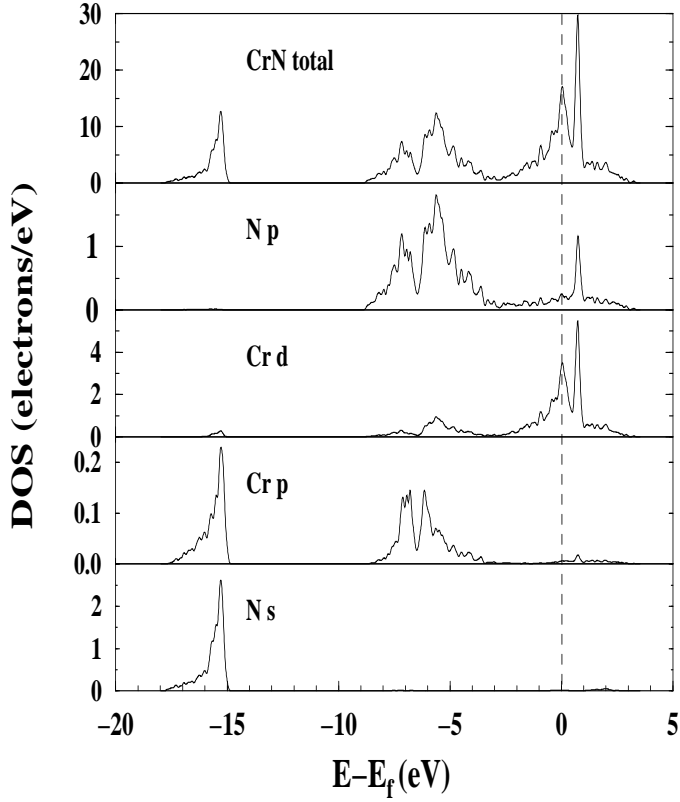}
\caption{
Density of states of cubic PM CrN. Top panel show the total DOS,
the panels below show the major contributions resolved by angular momentum
character.
The sharp peak at E$_F$ is entirely Cr $d$ (note that the N $p$
contribution is smooth through this region).
\label{dos_pm}}
\end{figure}

The stabilization due to the magnetic ordering can be understood from 
the analysis of the density of states (DOS). Figure \ref{dos_pm} shows
the PM CrN DOS for the cubic structure in orthorhombic symmetry
obtained from broadened eigenvalues from a regular mesh of
k points in the irreducible zone, which is sufficient for our purposes.
Unless otherwise indicated, in what follows all quantities refer to 
the theoretical lattice constant. 
As pointed out previously\cite{ppkb}, among the series of 3d TMNs,
CrN (as well as NaCl-type FeN \cite{sss}) is distinguished for the high DOS 
at E$_F$. Our DOS agrees well with the full-potential linearized 
augmented plane wave calculation of Ref.\onlinecite{ppkb}. At least 95\% of 
DOS at E$_F$ comes from Cr $d$ states, and the remaining 
contribution is due to the N $p$ states.
However, the DOS peak at E$_F$ is {\it entirely} due to Cr $d$ states, 
indicating that the peak arises from flat bands solely Cr in character.
Singh and Klein\cite{sk} have emphasized that this peak is Cr $t_{2g}$.
As we show below, it is not possible to identify this flat band solely
from symmetry line data.

\begin{figure}
\epsfxsize=8.5cm
\epsffile{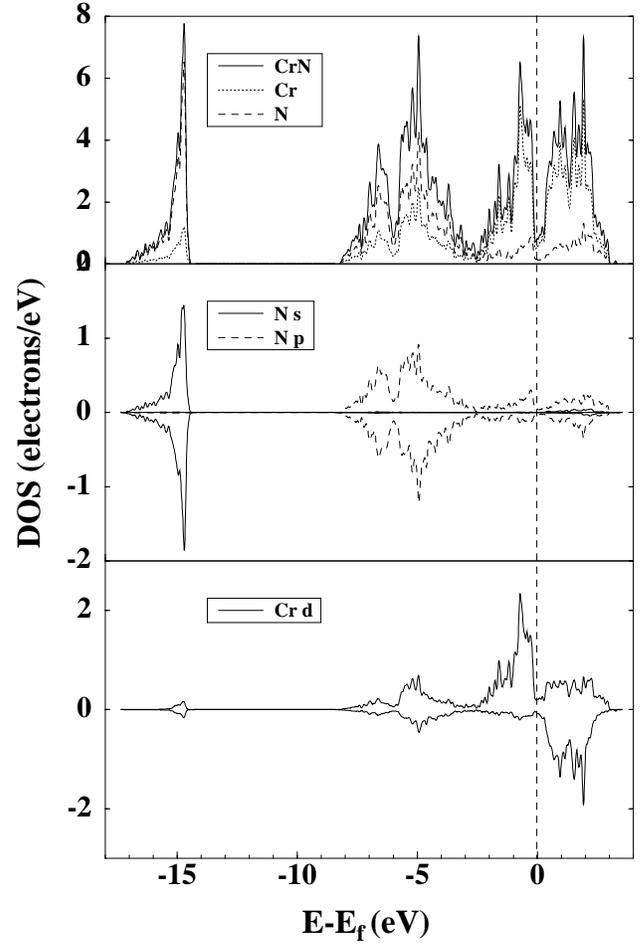}
\caption{
Density of states of cubic AFM CrN. Top panels show total and atom by atom
DOS, middle and bottom panels some contributions resolved by angular-moment.
\label{dos_afm}}
\end{figure}

Figure \ref{dos_afm} shows total and partial DOSs of the cubic
AFM$^2_{[110]}$ structure.
The peak at E$_F$ has disappeared: the AFM$^2_{[110]}$ ordering causes 
a large band splitting and the depletion (a pseudogap $\sim$ 0.5 eV wide) 
in the DOS of a sharp region around E$_F$.

Up- and down-polarized atoms are connected by translation plus 
spin-flip symmetry operations, thus contributions to DOS coming from 
the type of Cr and N atoms not shown in the Figure are equal to that
shown but with up and down components interchanged. For N (middle panel
of Figure \ref{dos_afm}) 
magnetic effects are negligible, although not zero by symmetry:
the magnetic moments are only $\sim$0.01 $\mu_B$. 
For Cr (bottom) the DOS around E$_F$ clearly resembles the total one,
hence the band splitting due to the polarization arises almost entirely 
from Cr $d$ states.

\subsection{Fermi Surface Properties}

\begin{figure}
\epsfxsize=7cm
\epsffile{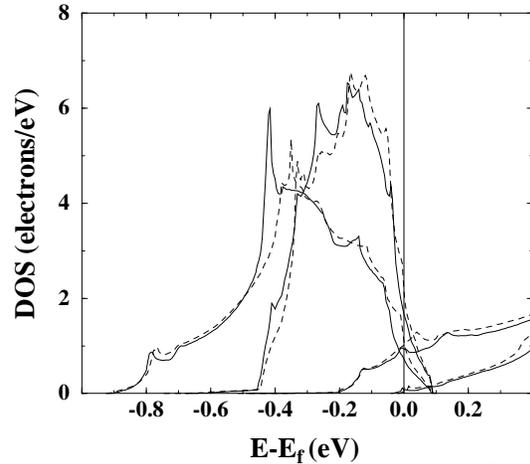}
\caption{
Band-by-band contribution of the four bands around the Fermi level, for both
cubic (dashed line) and distorted (full line) AFM$^2_{[110]}$ structure.
\label{dos_det}}
\end{figure}

Notwithstanding the strong stabilization by moving states away from E$_F$,
AFM$^2_{[110]}$ ordering does not manage to open a gap at the E$_F$, and 
the system is still
a (relatively low carrier density) metal, at variance 
with one\cite{hhvp} experimental characterization as a small gap 
semiconductor but in agreement with the results of Browne 
{\it et al.}\cite{blsm}
on very nearly stoichiometric samples.
To provide a clearer picture of our prediction, 
in Figure \ref{dos_det} we show
a band-resolved DOS including only the
four bands (see Section \ref{bsmo}) that cross or touch the Fermi level.  
These DOSs were calculated from a 528 k-points mesh, interpolated using 
a Fourier spline technique\cite{spline}, and evaluated using the linear 
tetrahedron method. 
Full lines refer to the distorted (experimental) structure AFM$^2_{[110]}$
bands, while dashed lines refer to the cubic AFM$^2_{[110]}$ phase. 
Basically three bands contribute to the DOS at E$_F$,
and a fourth band has its minimum essentially at E$_F$. 

In Table \ref{omega} the band-by-band values of the band filling,
density of states at E$_F$, and the Fermi velocities are given.  From
the totals we can evaluate the Drude plasma energies $\hbar \Omega_p$:
\begin{equation}
\Omega_{p,j}^2 = 4 \pi e^2 N(E_F) v_{F,j}^2,
\end{equation}
where $j$ denotes any of the Cartesian axes.  We obtain $\hbar \Omega_{p,x}$
= 2.3 eV, $\hbar \Omega_{p,y}$ = 2.0 eV, $\hbar \Omega_{p,z}$ = 2.6 eV,
so the non-cubic nature is evident but is only a $\pm$ 15\% effect.  This
difference is due to the antiferromagnetism, not to the structural
distortion whose effect we have found to be much smaller.  
These values of plasma energies are
rather small for transition metal compounds, but are typical of metallic
(not semiconducting or even semimetallic) behavior.

\subsection{Analysis of Magnetic Structures}

\begin{figure}
\epsfxsize=6cm
\epsffile{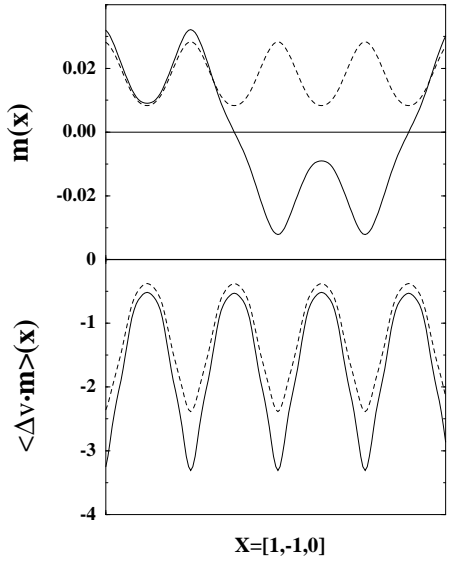}
\caption{
Top panel: the planar average of local magnetization (in units $\mu_B$/bohr)
is pictured along the $\hat{x}=[1\overline{1}0]$ direction of the orthorhombic
cell for FM (dashed line) and AFM$^2_{[110]}$ (full line) phases.
Bottom panel: planar averages (in Ryd$\cdot\mu_B$/bohr) of the
corresponding exchange-correlation potential splittings (see text).
\label{magn}}
\end{figure}

We mention a technical detail about calculation of magnetic 
moments. In case of an AFM structure with aligned spins along high 
symmetry directions, it is very suited for a  plane-wave basis to extract 
magnetic moments from the planar average of local magnetization,
which is a more direct and simple procedure than the common way of 
projecting plane waves onto atomic-like orbitals.
In the top panel of Figure \ref{magn} the planar average of local 
magnetization along $\hat{x}$ for the FM (dashed line) and AFM$^2_{[110]}$ 
(full line) phase is shown. Peaks are in correspondence of atomic positions, 
and the integrals over interlayer distances give by construction
the layer-by-layer magnetization, i.e. the magnetization per couple 
chromium-nitrogen, or, in practice, the magnetization per Cr atom,
being that of N atoms almost discardable. The idea to obtain contributions
to some physical quantity coming from individual space regions, not 
accessible from its macroscopic (or 'integrated') value, using planar 
and macroscopic averages has been sometime used. 
It leaded, for example, to introduce
quantities such as energy\cite{chma} and stress\cite{ff} density. 
In order to the integrals of the planar averages be well (uniquely) defined, 
a typical necessity is that the average is referred to an appropriate bulk 
symmetry plane (a critical case is, for instance, the average on (111) 
direction for a zincblebde structure: the atomic layers are not equally 
spaced and the volume attained to a single atom is undefined).
We return at Figure \ref{magn} below.

To complete the analysis of the DOS, in Figure \ref{dos_fm} we present 
results for the FM phase of CrN in the experimental structure. As expected, 
the total DOS at the E$_F$ is dominated by the Cr $d$ states.  
Ferromagnetism results in the splitting of 
up and down $d$ type bands,
and a strong stabilization with respect to the PM phase occurs.
The competition between antiferromagnetism and ferromagnetism for metals
is usually an intriguing question. In the present case, we can attribute 
the stronger AFM stability to a larger average spin splitting of Cr 
3$d$ states. It is clear from top panel of Figure \ref{magn} that
for AFM$^2_{[110]}$ there is larger magnetization on the Cr atoms. 
This trend should also be reflected in the band splitting. A measure of 
the average band splitting per atom is taken to be:

\begin{eqnarray}
\langle \Delta V_{xc}\rangle_m =
\int_{S_j} dr\,\Delta V_{xc}(r)\cdot m(r)/\int_{S_j} dr~m(r)
\label{eigsp}
\end{eqnarray}

\noindent
{\it i.e.} the up and down exchange-correlation potential difference 
$\Delta V_{xc}$, weighted by the local magnetization, integrated over 
the inter-planar volume $S_j$ and normalized to the atomic moment. 
The integrand of Equation \ref{eigsp} is 
pictured in the bottom panel of Figure \ref{magn}; the supralinear dependence
of Kohn-Sham potential on magnetization leads to the difference between
AFM$^2_{[110]}$ (full line) and FM (dashed) 
$\Delta V_{xc}({\bf r}) \cdot m({\bf r})$ 
in the regions around the atoms, resulting in an integrated value (over 
the interplanar distance) of 2.58 eV for the distorted AFM$^2_{[110]}$, 
2.52 for the cubic AFM$^2_{[110]}$, and 2.02 eV for the distorted FM phase. 
The calculated splitting roughly reflects the magnitude visible in the 
Figures of DOS previously discussed. In Figure \ref{diff} an expanded
view of the partial DOS of Cr around E$_F$ is shown. Evidently, the 
AFM$^2_{[110]}$ DOS has a minimun at E$_F$, whereas for the FM system E$_F$ 
falls on the edge of the second peak of Cr d states that is unoccupied 
in the PM phase (see Figure \ref{dos_pm}). 

\subsection{Band Shift from Magnetic Order and Lattice Distortion}
\label{bsmo}

The band splitting mechanism is further clarified by 
analysis of band shifts. Figures \ref{band_pm} and \ref{band_afc} 
show the PM and the AFM$^2_{[110]}$ band structures, respectively, 
calculated in the orthorhombic cell. 
The coordinates of the high-symmetry k-points are, in the orthorhombic
reference system (see bottom panel of Figure \ref{band_pm}), $M$=($\pi/a$,0,0),
$X$=(0,$\pi/b$,0), $L$=(0,$\pi/b$,$\pi/c$), and $W$=($\pi/a$,$3\pi/2b$,0),
where $a$, $b$, and $c$ are given in Section \ref{csmc}.
In the cubic PM phase, at points $M$, $X$, $W$, and $L$ all bands are doubly 
degenerate, due to the fcc to orthorhombic band downfolding. 
Symmetry breaking due to the AFM$^2_{[110]}$ ordering splits the degeneracy 
along the direction connecting points $M$ and $W$, whereas the lattice shear 
($\alpha\neq$ 90$^o$) causes the band splitting at $X$ and $L$. 
The former effect 
is the most effective in stabilizing the system. The splitting ($\sim$ 2 eV) 
strongly depletes the region around E$_F$, leaving only few band crossings.
It is particularly visible at $M$ by comparing cubic PM and AFM$^2_{[110]}$ 
band structures.

The effect of lattice shear produces a further stabilization, i.e. a 
further band opening, but is much less spectacular. It is emphasized
in Figure \ref{band_pm}: band energies shown along $M$-$\Gamma$-$X$ are 
those of the cubic PM phase, whereas along $X$-$W$-$L$ the Figure 
reports values for the distorted structure, so that the splitting is 
visible at $X$. This effect is very small (its order of magnitude 
is only some tens of meV), and band structures of cubic and distorted 
AFM$^2_{[110]}$ phase (Figures \ref{band_afc} and \ref{band_afd}) 
differ only in certain regions.

An unusual feature that can be seen in Figure \ref{band_afc} is the 
coincidence (or nearly so) of E$_F$ with four band crossings along
the directions shown: two along the $M$-$\Gamma$ line, one along 
$\Gamma$-$X$, and one at $W$. At general k-points away from these crossings, 
the bands will repel, hence they will not contribute to N(E$_F$), but rather 
to a pseudogap at E$_F$.

\section{Conclusion}
Our first principles results for CrN confirm most of the available
experimental data. The observed (110) double-sheet distorted structure
is found to be the lowest in energy, stable not only against paramagnetic and
ferromagnetic phases, but also with respect to other antiferromagnetic
arrangements. We found the system to be a weak metal, with large
magnetic moments on Cr atoms. 

Although the fundamental features 
favouring a certain spin arrangement with respect to another are
still to be understood, we are able to clarify the role played by
the distortion by disentangling its effect and that of the magnetic
ordering. Magnetic ordering is found to be dominant in determining both
structural and magnetic properties of the system, with respect to which
the distortion can be seen as a small perturbation. Nevertheless,
distortion appears decisive to resolve the competition among the different
antiferromagnetic phases considered here. Indeed, it causes
a meaningful energy gain for the (110) double-sheet phase, possibly 
due to relieving tensile stress stored in the spin-antiparallel 
metal-metal bonds.

Our calculations represent one of the first attempts to study
magnetic systems within a framework of plane waves with ultrasoft
pseudopotetials. All comparisons with previous available results
confirm that the accuracy of this methodology is comparable to that
of the all-electron calculations.

\section*{Acknowledgments}
This work was supported by National Science Foundation Grant No.
DMR-9802076. Computations were carried out at the San Diego 
Supercomputing Center and at the Maui High Performance Computing 
Center.

 
\begin{table}
\caption{$E_c^{th}$ denotes energies (in eV/formula unit) for cubic
structures calculated at
the theoretical $a_0$ of the respective phases; $E_c^{ex}$'s are relative
to the experimental $a_0$ for all phases. $E_d^{th}$'s and $E_d^{ex}$'s
are the same as before but now the structures are distorted by the observed
change of angle $\alpha$.
All energies are referred to the that of the most stable structure, i.e.
the AFM$^2_{[110]}$ phase with theoretical $a_0$ and distorted $\alpha$.
$m$'s are magnetic moments (in $\mu_B$/metal atom).
\label{energy}}
\begin{tabular}{cccccc}
  & AFM$^2_{[110]}$ & AFM$^1_{[110]}$ & AFM$_{[111]}$ & FM & PM \\
\hline
$E_c^{ex}$  & 0.068  & 0.034   &  0.085  & 0.110 & 0.273   \\
$E_c^{th}$  & 0.068  & 0.033   &         &       & 0.259    \\
\hline
$E_d^{ex}$  & 0.020  & 0.043   &         & 0.085 & 0.244    \\
$E_d^{th}$  & 0      &         &         &       & 0.235    \\
\hline
$m^c$       &  2.17  & 1.90    & 2.28    & 1.92  &          \\
$m^d$       &  2.15  & 1.94    &         &  2.08 &         \\
\end{tabular}
\end{table}
 
 
\begin{table}
\caption{
Band-by-band, and total values of the band filling (in electrons per
Cr$_4$N$_4$ cell) DOS at the Fermi
level (states/eV), and the rms values of the
Cartesian components of the Fermi
velocity (10$^8$ cm/s).  The last line gives the total electrons and
DOS, and the rms velocities at E$_F$ for all four bands.
\label{omega}}
\begin{tabular}{cccccc}
Band  & Occupation & N(E$_F$) & v$_x$ & v$_y$ & v$_z$  \\
\hline
1    &  1.97      & 0.73    & 0.084   & 0.056  &  0.061  \\
2    &  1.93      & 1.67    & 0.10    & 0.063  &  0.036  \\
3    &  0.10      & 0.93    & 0.16    & 0.16   &  0.25   \\
4    &  0.002     & 0.08    & 0.12    & 0.16   &  0.074  \\
\hline
Total & 4.00      & 3.41    & 0.12    & 0.10   &  0.14  \\
\end{tabular}
\end{table}
 
 
\begin{figure}
\epsfxsize=7cm
\epsffile{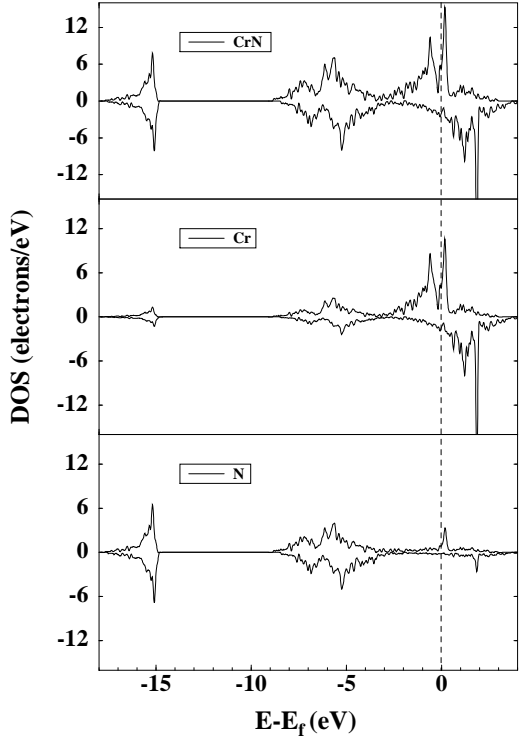}
\caption{
Density of states of FM CrN calculated in the experimental structure.
Top panel is the total DOS, lower panels show the individual contributions
of each atom, as indicated in the Figure.
\label{dos_fm}}
\end{figure}
 
 
\begin{figure}
\epsfxsize=7cm
\epsffile{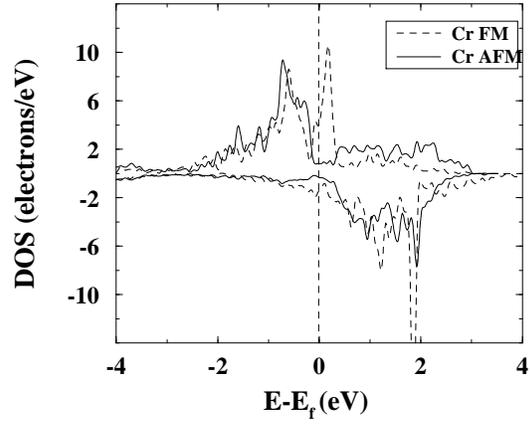}
\caption{
Contributions of Cr to total DOS: comparison between AFM$^2_{[110]}$ and
FM results.
\label{diff}}
\end{figure}
 
 
\begin{figure}
\epsfxsize=8.5cm
\epsffile{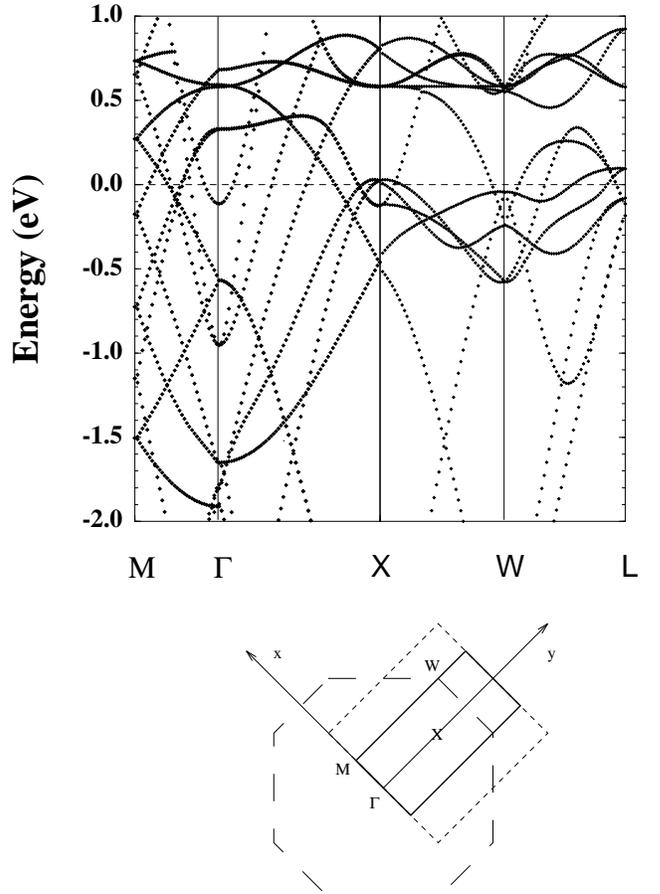}
\caption{
Top: band structure for cubic PM CrN in the orthorhombic BZ.
Energies between points $M$ and $X$ are for the cubic structure,
where between $X$ and $L$ are for the distorted one, so that at $X$
the minor band splitting due to the distortion is visible.
Bottom: high symmetry points of the CrN BZ. Full line denotes the
orthorhombic BZ, dashed line the cubic BZ and long-dashed line the fcc
truncated octhaedron.}
\label{band_pm}
\end{figure}
 
 
\begin{figure}
\epsfxsize=8.5cm
\epsffile{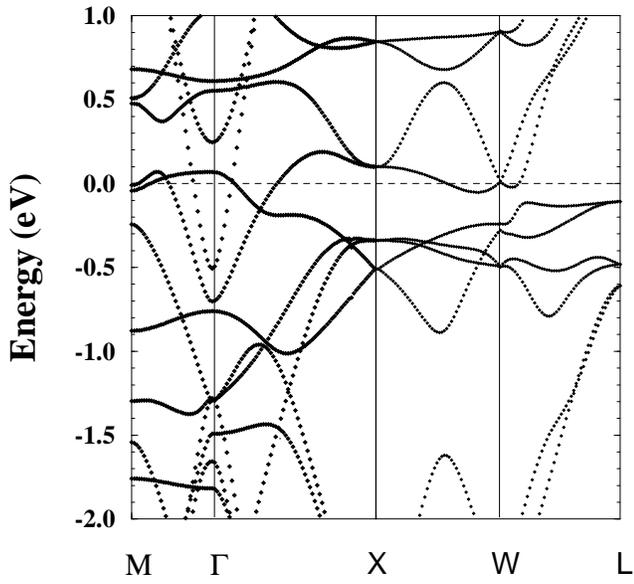}
\caption{
Band structure for the cubic AFM$^2_{[110]}$ CrN. At variance with the PM
structure, the double degeneracy at points $M$ and $W$ due to the
cubic-to-orthorhombic downfolding is broken by the AFM$^2_{[110]}$ symmetry,
thus allowing the (mostly Cr d-like) bands to be shifted out from the E$_F$.}
\label{band_afc}
\end{figure}
 
 
\begin{figure}
\epsfxsize=8.5cm
\epsffile{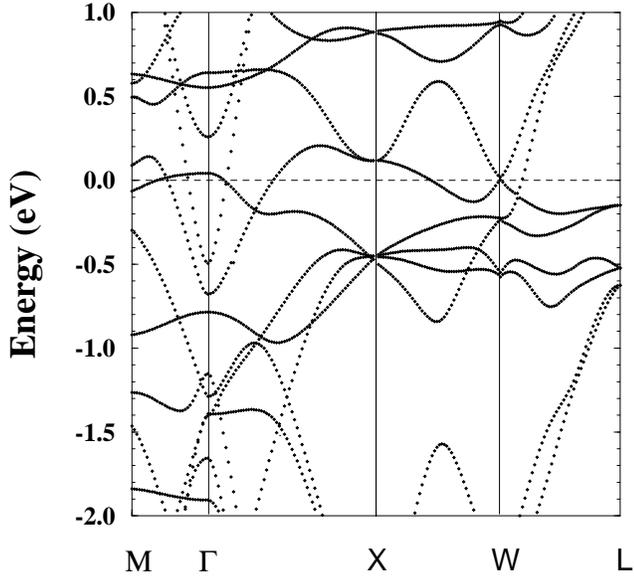}
\caption{
Band structure for the distorted AFM$^2_{[110]}$ CrN. Lifting of degeneracies
can be observed, but there are no dramatic changes with respect to the
cubic AFM$^2_{[110]}$ structure.
\label{band_afd}}
\end{figure}

\end{document}